\documentclass[prc,aps,amsfonts,amsmath,superscriptaddress,twocolumn,showpacs,floatfix]{revtex4-1}
\usepackage{graphicx} 
\usepackage{epsfig} 

\begin{document}

\title{Localization and clustering in the nuclear Fermi liquid}

\author{J.-P. Ebran}
\affiliation{CEA/DAM/DIF, F-91297 Arpajon, France}
\author{E. Khan}
\affiliation{Institut de Physique Nucl\'eaire, Universit\'e Paris-Sud, IN2P3-CNRS, F-91406 Orsay Cedex, France}
\author{T. Nik\v si\' c}
\author{D. Vretenar}
\affiliation{Physics Department, Faculty of Science, University of
Zagreb, 10000 Zagreb, Croatia}

\begin{abstract}
Using the framework of nuclear energy density functionals we examine
the conditions for single-nucleon localization and formation of 
cluster structures in finite nuclei. We propose to characterize
localization by the ratio of the dispersion of single-nucleon wave
functions to the average inter-nucleon distance. This parameter 
generally increases with mass and describes the gradual transition 
from a hybrid phase in light nuclei, characterized by the spatial
localization of individual nucleon states that leads to the formation
of cluster structures, toward the Fermi liquid phase in heavier
nuclei. Values of the localization parameter that correspond to a
crystal phase cannot occur in finite nuclei. Typical length and energy
scales in nuclei allow the formation of liquid drops, clusters, and
halo structures. 
\end{abstract}


\date{\today}

\maketitle

Nucleons in atomic nuclei and extended nuclear matter exhibit a
variety of phases. Liquid drop aspects, for instance, were first
inferred \cite{BW} from fission properties in heavy nuclei. Soon 
afterwards it was also predicted and observed that cluster states 
could occur, especially in light nuclei \cite{4}. Halo structures
in nuclei were discovered in the late 1980's \cite{halo}. Although
a number of theoretical models have been developed that successfully
describe particular features of these nucleonic phases, open
questions remain: can nucleon crystal states occur?; do all
nucleonic phases (liquid, cluster, halo, crystal) have a common origin
and, therefore, can they be described in a unified theoretical
framework? In particular, in a recent study \cite{nat} we have shown that the
confining nuclear potential determines the degree of localization and 
clustering in finite nuclei. In the present work we analyze the 
emergence of cluster states, considered as
a transitional phase between a quantum liquid (nuclear matter) and a
solid (crystal). These considerations are also relevant
for the description of the crust of neutron stars, where it is known
that decreasing matter density (further from the center of the star)
leads to a transition from the nuclear matter phase (liquid) to a 
Wigner crystal, with a pasta (cluster) phase in between
\cite{bay,lat,wat}.

Clustering -- the arrangement of nucleons in clusters of bosonic 
characters, especially in light nuclei, coexists with the nuclear
mean-field. The nature of the cluster phase itself is very much under
debate: can nuclei in the cluster phase behave like a dilute gas of
$\alpha$-particles? \cite{toh01,zin08,fun09} This refers to the 
localization of the $\alpha$'s with respect to the size of the
nucleus.  Here we firstly address the question of localization of
nucleons: what is the mechanism of confinement of individual nucleons
into clusters such as, for instance, $\alpha$-particles? Since the
majority of theoretical approaches that quantitatively describe cluster
states assume {\em a priori} the existence of such structures (or
facilitate their formation by employing Gaussian wave functions 
centered at given positions in space), and the corresponding effective
interactions are adjusted to the binding energies and scattering phase
shifts of these configurations, one cannot say that the initial
localization of nucleons and the mechanism that drives the transition
from the fermionic liquid to cluster structures are fully understood
\cite{MF-Nature}. As shown in our previous study \cite{nat}, there is
a direct correlation between the effective potential that
confines the neutrons and protons to the nucleus, and the enhancement
of the symmetries of the clustering. The deformation of the nucleus
also contributes to the formation of clusters because it removes the
degeneracy of single-nucleon levels associated with spherical symmetry
\cite{4}. Clustering effects are, of course, more dominant in excited
nuclear states, and this can be understood from the fact that the
closeness to the particle emission threshold favors cluster formation.
States close to the continuum cannot be isolated from the environment
of scattering states, so cluster states at the threshold belong to an
open quantum system \cite{mar}. The origin of cluster formation,
however, lies in the effective nuclear interaction, and a fully
microscopic description of clustering necessitates a framework that
encompasses both cluster and quantum liquid-drop aspects in light and
heavier nuclei \cite{nat,rein}.

The issue of solid (crystal) vs. quantum liquid nature of nuclei was
already addressed by B. Mottelson, who emphasized that the 
essence of the concept of independent particle motion is the 
fact that the orbits of individual nucleons are delocalized 
and reflect the shape and radial dependence of the effective 
potential over the entire nucleus \cite{28}. Mottelson used the 
quantality parameter \cite{boe}:
\begin{equation}
\label{lambda}
\Lambda\hat{=}\frac{\hbar^2}{m{\bar r}^2V'_0} \; ,
\end{equation}
with the strength of the bare nucleon-nucleon interaction
$V'_0 \sim 100$ MeV, and the inter-nucleon equilibrium distance $\bar r$, 
to characterise the transition between quantum liquid and crystalline 
solid phases. The quantality $\Lambda$ is defined as the ratio of the 
zero-point kinetic energy of the confined particle to its potential energy, 
and the transition occurs in the region $\Lambda \simeq 0.1$. The 
typical value obtained for nuclear matter ($m$ being the nucleon mass): 
$\Lambda \simeq 0.5$, is characteristic for a quantum liquid phase \cite{28}. 
However, the parameter $\Lambda$ is defined for infinite 
homogeneous systems and its applicability to finite nuclei  
is limited by the fact that it does not include any nuclear mass or size dependence.  
Cluster states in finite nuclei introduce an additional
phase of nucleonic matter. In fact, if instead of the 
nucleon-nucleon potential one considers an 
alpha-alpha potential \cite{mich98} for V'$_0$ in Eq.~(\ref{lambda}) , 
the value of the quantality parameter decreases to $\Lambda \simeq$ 0.1,
entering into the liquid to crystal phase transition region \cite{zin07}. 

To analyze localization and the occurrence of clustering in finite nuclei 
we need to consider a quantity that is sensitive to the nucleon number 
and size of the nucleus. Two characteristic lengths 
quantify the crystalline vs. Fermi liquid transition, similar to 
the condensed matter case \cite{29}: the localization of the constituent 
wave functions in the system, and the average inter-constituent distance. 
Hence the localization of the single-nucleon wave function, and 
eventually the degree of nucleonic density clustering, 
can be quantified by the dimensionless parameter $\alpha$ introduced 
in Ref.~\cite{nat}: 
\begin{equation}\label{eq:adef}
\alpha \hat{=}\frac{ \Delta r}{\bar r}
\end{equation}
where $\bar r$ is the average inter-nucleon distance, and $\Delta r$
the spatial dispersion of the wave function: 
\begin{equation}\label{eq:sig}
\Delta r=\sqrt{\left<r^2\right>-\left<r\right>^2}
\end{equation}
We propose to use the parameter $\alpha$ to study localization effects
in nuclei. For large values of $\alpha$ the orbits of individual
nucleons will be delocalized and the nucleus in the Fermi liquid
phase.  On the other hand, when $\alpha$ is small nucleons will 
be localized on the nodes of a crystal-like structure. At intermediate
values one expects a transition from the quantum liquid phase to a
hybrid phase of cluster states. For finite systems like nuclei this
transition, of course, cannot be sharp. In a first approximation one
expects that the transition occurs for $\alpha \approx 1$ because for 
this value the spatial dispersion of the single-nucleon wave function
is of the same size as the inter-nucleon distance and, therefore, optimal 
for nucleons to form a correlated cluster such as an alpha particle.

Localization parameters have also been considered for other quantum 
systems, such as quantum dots \cite{yan99}, or in condensed 
matter \cite{fal05}, to characterize the occurrence of a hybrid phase 
between the liquid and crystal phases. However, in general it will 
not be possible to find a universal and quantitative localization 
parameter that can be applied to different quantum systems, 
because the transition from the quantum liquid to the
crystal phase is controlled by the specific dynamics and 
length scale of the system under consideration \cite{yan99,fal05}.
In the nuclear case, in particular, finite size effects are important. 
It will be shown that the parameter $\alpha$ can be used to 
qualitatively characterize transitions between different phases 
of nucleonic matter.

In a first, non self-consistent, approximation one can use a 
3-dimensional isotropic harmonic oscillator (HO) for the confining
nuclear potential.  This approximation allows for a qualitative
discussion of the effects of the effective nuclear interaction on the
spatial arrangement of nucleons.  The 3-dimensional HO wave functions
$\varphi_{klm}(\vec{r})$ for the first $s$, $p$ and $d$ states, which
provide the main contribution to cluster states in light nuclei, read
\cite{27}:
\begin{equation}\label{eq:wf}
\varphi_{0lm}(\vec{r}) \sim
\frac{r^l}{b^{(3/2+l)}}e^{-\frac{r^2}{2b^2}}Y_l^m(\hat{r})\; ,
\end{equation}
where b is the oscillator length defined by:
\begin{equation}\label{eq:b}
b\hat{=}\sqrt{\frac{\hbar}{m\omega_0}}=\frac{\sqrt{\hbar R}}{\left(2m V_0\right)^{1/4}} \;.
\end{equation}
R is the radius of the potential for which $V=0$, and $V_0$ denotes
the depth of the potential at $r=0$. It should be emphasized that charge
radii of atomic nuclei are determined with high precision in electron
scattering experiments, in contrast to the depth of a confining
potential $V_0$ which is experimentally poorly constrained.

A straightforward calculation yields the spatial dispersion $\Delta r \approx 0.5 b$ for the
first $s$, $p$ and $d$ HO wave functions, and they display a Gaussian-like radial
dependence (Eq.~(\ref{eq:wf})). Consequently, for a constant radius R,
a deeper potential V$_0$ implies a smaller value of the oscillator
length $b$ (Eq.~(\ref{eq:b})), and thus a smaller dispersion. This
concept can be extended to the more general case of deformed nuclei by
approximating the confining potential with an axially deformed HO. The
wave functions are then expressed as \cite{26,aga}
\begin{widetext} 
\begin{equation}\label{eq:wfd}
\varphi_{n_r,n_z,m_l}(r,\phi,z) \sim e^{im_l}\left(\frac{r}{b_\bot}\right)^{m_l}H_{n_z}(z/b_z)
L^{m_l}_{n_r}(r^2/b^2_\bot)e^{-\frac{1}{2}\left(\frac{z^2}{b_z^2}+\frac{r^2}{b_\bot^2}\right)}\; ,
\end{equation}
\end{widetext}
where H and L are the Hermite and Laguerre polynomials, respectively.
Eq.~(\ref{eq:wfd}) displays a radial dependence similar to the 3D
isotropic case (Eq.~(\ref{eq:wf})): the dispersion of the wave
functions depends on the oscillator lengths b$_z$ and b$_\bot$ in the
respective directions, which in turn depend on the depth of the
potential. In the deformed HO approach, the depth of the potential,
therefore, determines the localization of nucleon wave functions, just
like in the spherical case.

\begin{figure}[tb]
\begin{center}
\scalebox{0.35}{\includegraphics{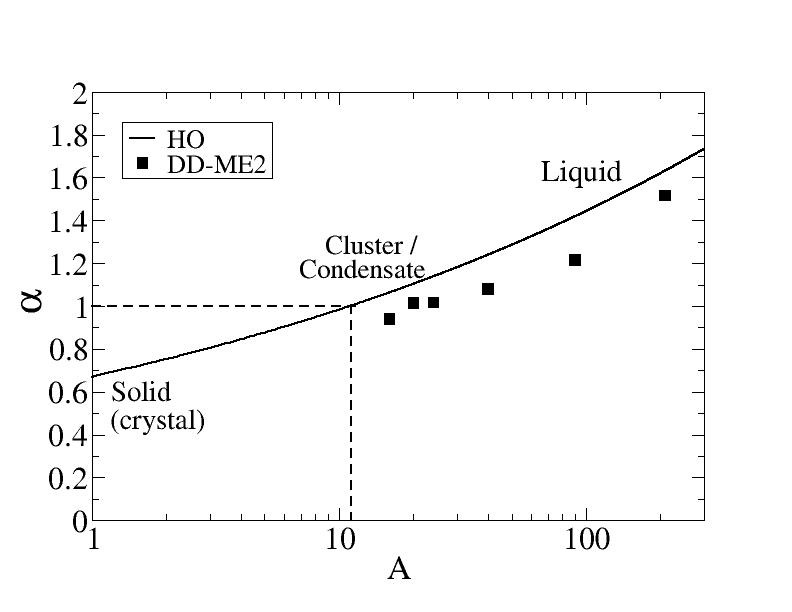}}
\caption{The localization parameter $\alpha$ (Eq. (\ref{eq:alpha2})) 
as a function of the number of nucleons. The average values of $\alpha$ for $^{16}$O 
$^{20}$Ne, $^{24}$Mg, $^{40}$Ca, $^{90}$Zr, calculated for the 
microscopic self-consistent solutions obtained using the functional DD-ME2, 
are denoted by squares.}
\label{fig:alpha}
\end{center}
\end{figure}

The localization parameter $\alpha$ obtained using expression (\ref{eq:b}) for
the harmonic oscillator length reads:
\begin{equation}\label{eq:alpha}
\alpha\simeq\frac{b}{r_0} = \frac{\sqrt{\hbar R}}{r_0(2mV_0)^{1/4}}\;, 
\end{equation}
with $r_0 = 1.25$ fm \cite{boh}. Using the liquid drop
parameterization for the radius R=r$_0$A$^{1/3}$, Eq. (\ref{eq:alpha})
reads
\begin{equation}\label{eq:alpha2}
\alpha=\frac{\sqrt{\hbar}A^{1/6}}{(2mV_0r_0^2)^{1/4}}\simeq 0.67 A^{1/6} \;.
\end{equation}
Figure \ref{fig:alpha} displays the evolution of $\alpha$ 
with A, for a typical values of V$_0$= 70 MeV. 
The localization parameter $\alpha$ generally increases with the 
number of nucleons and, therefore, cluster states are more easily 
formed in light nuclei, as observed experimentally \cite{4}.
The transition from localized clusters to a liquid state 
occurs for nuclei with $A \approx 30$. For heavier systems $\alpha$ is
considerably larger than 1 and, therefore, heavy nuclei consist of
largely delocalized nucleons and this explain their liquid drop nature
and the large mean free path of nucleons. More precisely,
nuclei are in the Fermi liquid phase and localized cluster states 
(hybrid phase) can be formed in light nuclei. 
Fig.~\ref{fig:alpha} also nicely
illustrates the fact that a crystal phase ($\alpha \lesssim  0.8$) cannot
occur in finite nuclei. However, nature may offer the possibility of
existence of nucleonic crystals in the crust of neutron stars, where
crystallization is caused by the long range Coulomb interaction in a
gravitationally constrained environment \cite{lat}. The transition
between the Wigner crystal and the quantum liquid in the neutron star
crust can be described by various models: gelification \cite{sator},
Coulombic frustration \cite{na07} or quantum melting \cite{fal05}.

In a fully microscopic analysis, the first two columns of
Table \ref{tab} display the values of the localization parameter
$\alpha$, calculated from Eq.~(\ref{eq:adef}) using the 
self-consistent ground-state solutions for the $N=Z$ nuclei $^{20}$Ne,
$^{24}$Mg, $^{28}$Si, $^{32}$S, and also the heavy $^{208}$Pb nucleus,
obtained with the functionals SLy4 \cite{22} and DD-ME2 \cite{23}.
These two functionals are representative of the two standard classes
of nuclear energy density functionals (EDFs): the non-relativistic and
relativistic functionals, and in Ref.~\cite{nat} they were used to
calculate the self-consistent equilibrium mean-field solution for
$^{20}$Ne. Both functionals reproduce the empirical ground-state
properties (binding energy, charge radius, matter radius) with a
typical accuracy of 1\%, as well as the quadrupole deformation of the
equilibrium shape. However, as it will be also shown in this study,
the density calculated with SLy4 displays a smooth behavior
characteristic of a Fermi liquid, whereas the functional DD-ME2
predicts an equilibrium density that is much more localized, with
pronounced cluster structures. The dispersions $\Delta r$ correspond
to the self-consistent single-nucleon Nilsson state [$1~1~0~1/2$],
which gives a pronounced contribution to clustering in these nuclei
\cite{nat}. Taking $\bar r = 0.9$ fm as a characteristic inter-nucleon
equilibrium distance \cite{28}, we determine the corresponding values
of the cluster parameter $\alpha$ (Eq. (\ref{eq:adef})), displayed in
the first two columns of Table \ref{tab}. In the four lighter nuclei
$^{20}$Ne, $^{24}$Mg, $^{28}$Si, $^{32}$S the $\alpha$ values
calculated with DD-ME2 are systematically smaller than those obtained
using SLy4, reflecting the more pronounced localization of the
nucleonic densities that was already observed in our previous study in
Ref.~\cite{nat}. While for light nuclei $\alpha \leq 1$, in the case
of $^{208}$Pb $\alpha$ is considerably larger than 1 and this
unambiguously characterizes the quantum liquid phase of nucleonic
matter in this nucleus. Note that for $^{28}$Si which is oblate in the
equilibrium state, the dispersion is calculated for the Nilsson state
[$1~0~1~1/2$]. We have verified that similar values are obtained for
other single nucleons states that build the cluster structures in
these nuclei and, also, that the localization parameter $\alpha$
averaged over all occupied states increases with mass number.

For completeness, in the last two columns of Table \ref{tab} we also 
list the values of the localization parameter $\alpha$ obtained using 
the HO expression Eq. (\ref{eq:alpha}) for the dispersion. In this calculation,
however, the nuclear radius $R$ and the depth of the potential $V_0$
are determined microscopically using the self-consistent equilibrium
solutions calculated with the EDFs SLy4 and DD-ME2. The trend is
similar to that obtained in the fully microscopic calculation, that
is, DD-ME2 predicts systematically smaller values of the localization
parameter $\alpha$. Note that this conclusion holds even when we
replace the nucleon bare mass in the denominator of Eq.
(\ref{eq:alpha}) with the effective mass $m^*$. The effective nucleon
mass for the functional SLy4 is $0.70 m$, and for DD-ME2 $0.66 m$. In
this case the value of the parameter $\alpha$ increases by a factor
$(m/m^*)^{1/4}\simeq1.1$, but the ratio between values that correspond to SLy4
and DD-ME2 is not altered by more than 1\%. 
\begin{table}[t]
\renewcommand{\arraystretch}{1.5}
\centering
\begin{tabular}{c|c|c|c|c|}
 \cline{2-5}
        & \multicolumn{2}{c|}{Self-cons.} & \multicolumn{2}{c|}{HO+EDF} \\
 \cline{2-5}
   & SLy4        & DDME2       & SLy4      & DDME2  \\
   \hline
$^{20}$Ne  & 0.99  &  0.97      & 1.00        & 0.97           \\
$^{24}$Mg  & 1.00  &  0.95     & 1.02        & 0.96           \\
$^{28}$Si  & 0.99  &  0.96     & 1.05        & 1.00           \\
$^{32}$S   & 0.99  &  0.96    & 1.06        & 0.99           \\
$^{208}$Pb & 1.28  &  1.31    & 1.46        & 1.40           \\
\hline
\end{tabular}
\caption{Left: the localization parameter $\alpha$ (Eqs. (\ref{eq:adef})) 
calculated from the fully self-consistent equilibrium solutions obtained 
with the EDFs SLy4 \cite{22} and DD-ME2 \cite{23}. Right: the same but using 
the 3D harmonic oscillator approximation (Eq. (\ref{eq:alpha})), for which 
SLy4 and DDME2 are used to determine the corresponding nuclear radii and depth of 
the confining potentials (see text).} 
\label{tab} 
\end{table}

In addition to the localization parameter obtained from the HO 
length (Eq. (\ref{eq:alpha2})), in Fig.~\ref{fig:alpha} we have also included 
the average values of $\alpha$ for $^{16}$O (0.94), 
$^{20}$Ne (1.02), $^{24}$Mg (1.02), $^{40}$Ca (1.08), $^{90}$Zr (1.22), 
calculated fully microscopically using the functional DD-ME2. These 
values are obtained by averaging the microscopic dispersions 
Eq.~(\ref{eq:adef}) for all occupied proton and neutron orbitals in
the self-consistent ground-state solution, and dividing by the  
characteristic inter-nucleon equilibrium distance $\bar r = 0.9$ fm \cite{28}. 
The microscopic average localization parameter describes the 
gradual transition from the hybrid phase, characterized by the 
spatial localization of individual nucleons, toward the Fermi liquid phase 
in heavier nuclei.

\begin{figure}[]
\vspace{5mm}
\scalebox{0.325}{\includegraphics{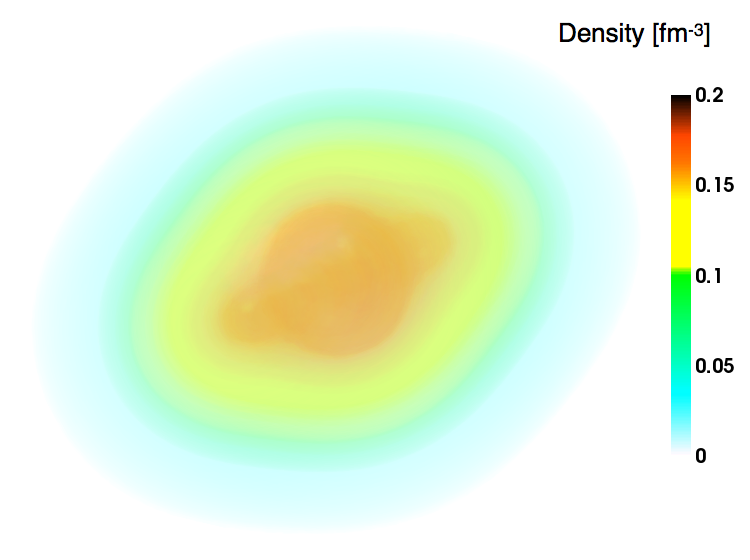}}\\
\scalebox{0.325}{\includegraphics{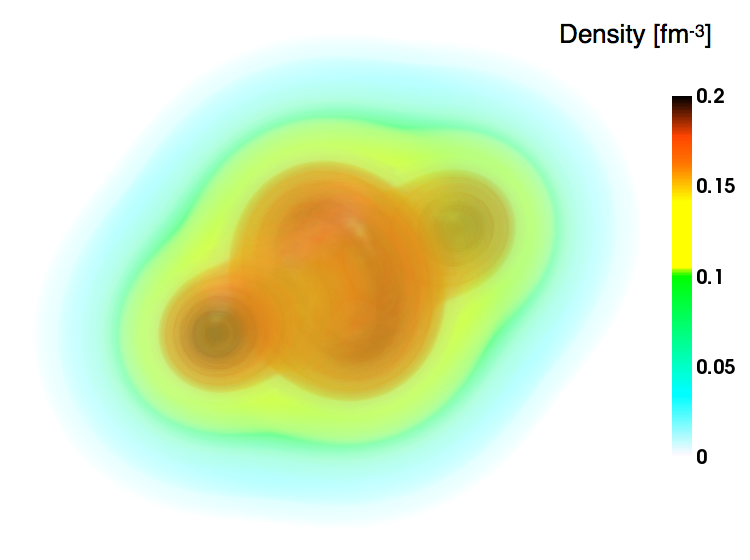}}
\caption{(Color online) Self-consistent ground-state densities of $^{20}$Ne, calculated with the 
energy density functionals SLy4 (top) and DD-ME2 (bottom). 
The densities (in units of fm$^{-3}$) are plotted in the 
the intrinsic frame of reference that coincides with the principal axes of the nucleus.}
\label{fig:clusters_Ne}
\end{figure}

\begin{figure}[]
\vspace{5mm}
\scalebox{0.325}{\includegraphics{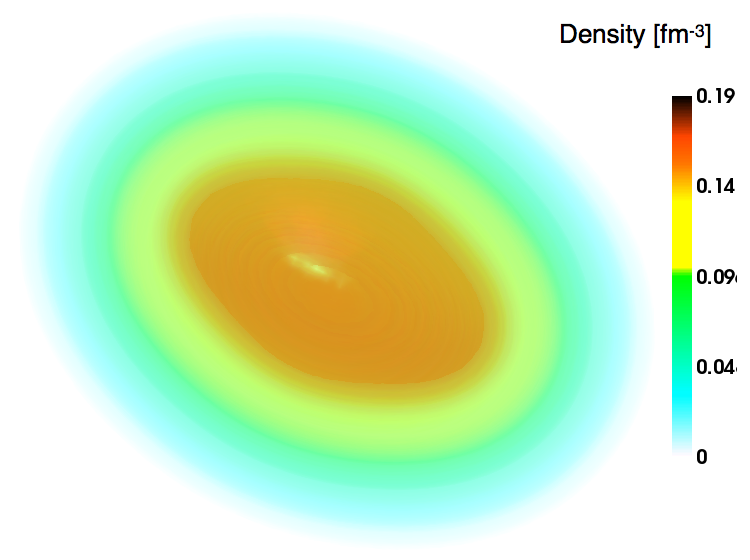}}\\
\scalebox{0.325}{\includegraphics{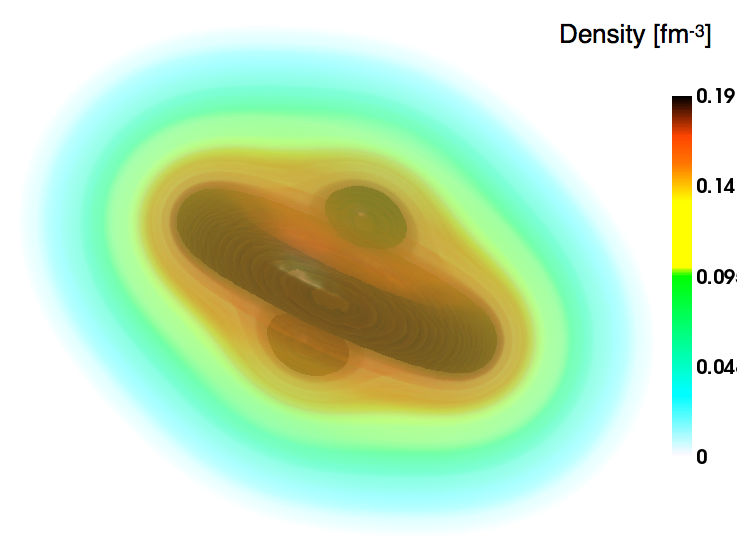}}
\caption{(Color online) Same as described in the caption to Fig. \ref{fig:clusters_Ne} 
but for the nucleus $^{28}$Si.}
\label{fig:clusters_Si}
\end{figure}

The localization of single-nucleon states can be analyzed in the HO
approximation but, of course, a harmonic oscillator potential cannot
give rise to clustering. Energy density functionals, on the other
hand, implicitly include many-body short- and long-range correlations
through their explicit density dependence and, therefore, should allow
formation of cluster-like substructures \cite{rein}. Most modern EDFs,
for instance, reproduce the binding energy and size of the $\alpha$
particle even though their parameters are not specifically adjusted to
very light nuclei. The different localization properties predicted by
the functionals SLy4 and DD-ME2 are reflected in the corresponding
nucleon density distributions. In Fig.~\ref{fig:clusters_Ne} we
display the corresponding axially and reflection symmetric
self-consistent equilibrium nucleon density distributions of
$^{20}$Ne. Although these functionals predict similar values for the
binding energy, charge and matter radii, and quadrupole deformation,
the corresponding equilibrium density distributions are rather
different. SLy4 yields a simple axially deformed prolate ellipsoid,
with only a slight indication of possible cluster formation. DD-ME2,
on the other hand, predicts two regions of pronounced localization at
the outer ends of the symmetry axis and an oblate deformed core. The
sharper density peaks will, of course, greatly enhance the probability
of formation of $\alpha$-clusters in excited states. We note that a
similar quasimolecular $\alpha$-$^{12}$C-$\alpha$ structure, although
with somewhat less pronounced clustering, was also obtained in the
Hartree-Fock calculation of Ref.~\cite{rein}, using the SkI3 Skyrme
functional. Since the nucleon effective masses for the two functionals
are very similar ($0.70 m$ for SLy4 and $0.66 m$ for DD-ME2),
the different level of localization and clustering predicted by SLy4
and DD-ME2 is partly related to the depth of the corresponding
confining Kohn-Sham potentials \cite{nat}, similar to the HO case (Eq.
(\ref{eq:alpha})). In Fig.~\ref{fig:clusters_Si} we show another
example, the equilibrium density distributions of $^{28}$Si calculated
with SLy4 and DD-ME2. In this case the equilibrium shape is oblate
($\beta \approx -0.35$), and again we find that DD-ME2 predicts the
formation of cluster-like structures, whereas nucleonic density shows
a smooth gradual decrease from the center of the oblate ellipsoid
calculated with SLy4. 

The present discussion can qualitatively be related to delocalized
wave functions of halo states in light nuclei \cite{han}. Several
subtle effects are at work in halo structures, such as the inversion
between the $p$ and $s$ orbitals, and the coupling to the continuum
\cite{ike10}. Here we only examine the delocalization of
single-particle wave functions. When the confining nuclear potential
is approximated by a square-well, the oscillations of the wave
function of a state of energy $E$ ($<$0) are determined by the wave
number $\hbar^2k^2=2m(E+V_0)$, whereas outside of the potential the
decay of the wave function is governed by $\hbar^2k'^2=-2mE$,
favoring a large radial extension for weakly bound states. However,
inside the potential a larger $k$ favors localization, and this occurs 
when the potential is deeper and/or $E$ gets
closer to zero. The degree of localization depends on the difference
$E-(-V_0)$, and a deep potential favors the localization of the wave
function for the spatial region located inside the potential. Also a
weakly-bound state will be more localized, and this means that the
confinement of nucleons into clusters is more likely to occur for
weakly-bound states close to the emission threshold. This in agreement
with the Ikeda picture, as well as subsequent studies
\cite{ike,4,toh01}. Therefore, a common feature of localized 
states (clusters) and haloes is that the energy of a state determines 
its spatial behavior either inside the potential (clusters)
or outside the range of the effective potential (halo states).

An important characteristic of nuclei is that
quantitatively the dispersion of the single-nucleon wave function can
be of the same order of magnitude as the inter-nucleon distance,
leading to clustering as discussed above. More generally, the typical
values of length scales and energies allow the formation of liquid
drops, clusters, and haloes in nuclei, but not crystals. 

\bigskip
This work was supported by the Institut Universitaire de France and by
the Croatian Ministry of Science, project 1191005-1010.

\end{document}